\documentclass[aps,prb,twocolumn,groupedaddress]{revtex4}

\preprint{H.Itoh, APL}
\draft
\usepackage{graphicx}
\usepackage{times, mathptmx}
\begin{document}

\title{Polarization-dependent photoluminescence-excitation spectra of
one-dimensional exciton and continuum states in T-shaped quantum
wires\footnote{to be published in Applied Physics Letters}}

\author{Hirotake Itoh}
\email[E-mail(H.Itoh):]{hiroitoh@issp.u-tokyo.ac.jp}
\author{Yuhei Hayamizu}
\author{Masahiro Yoshita}
\altaffiliation{also at Bell Laboratories, Lucent Technologies, 600 Mountain Avenue, Murray Hill, NJ 07974, USA}
\author{Hidefumi Akiyama}
\altaffiliation{also at Bell Laboratories, Lucent Technologies, 600 Mountain Avenue, Murray Hill, NJ 07974, USA}
\affiliation{Institute for Solid State Physics, University of Tokyo, and CREST, JST, \\
5-1-5 Kashiwanoha, Kashiwa, Chiba 277-8581, Japan}

\author{Loren N. Pfeiffer}
\author{Ken W. West}
\affiliation{Bell Laboratories, Lucent Technologies, 600 Mountain Avenue, Murray Hill, NJ 07974, USA}

\author{Marzena H. Szymanska}
\author{Peter B. Littlewood}
\altaffiliation{also at Bell Laboratories, Lucent Technologies, 600 Mountain Avenue, Murray Hill, NJ 07974, USA}
\affiliation{TCM group, Cavendish Laboratory, University of Cambridge, Cambridge CB3 0HE, UK}

\date{\today}

\begin{abstract}
We measured polarization-dependent photoluminescence-excitation spectra of highly uniform T-shaped quantum wires at 5 K.
We attribute one peak to the one-dimensional(1D)-exciton ground state and the continuous absorption band to 1D continuum states.
These had similar polarization dependences. We also observed some other peaks above the 1D-exciton ground state and attribute them to exciton
 states consisting of excited hole subbands. These results show good agreement with a model calculation of a single electron-hole pair in T-shaped
 geometry with exact diagonalizations of the Coulomb interaction.
\end{abstract}


\maketitle

In quantum wires, one-dimensional (1D) excitons are expected to have a large binding
energy~\cite{loudon59,elio60,ogawa_ContinuumSurpress_prb_91_rapid,ogawa_ContinuumSurpress_prb_91,wegscheider_firstLasing_93,someya_LaterallySqueezed_95,rossi_scaling_prl_97,goldoni_remote_prl_98,thilagam_jap_97,glutsch_ExcitonsinTwire_97,brinkmann_rydbelg_prb_97,walck_ExcitonEb_prb_97,zhang_Scaling_prb_99,stopa_BandgapRenormal_prb_01,szymanska_Exciton_withCoulomb_prb_01}
and large oscillator strength in the ground
state~\cite{loudon59,elio60,ogawa_ContinuumSurpress_prb_91_rapid,ogawa_ContinuumSurpress_prb_91,akiyama_PLEold_96}, while the oscillator strength of
exciton excited states and 1D continuum states should be suppressed~\cite{ogawa_ContinuumSurpress_prb_91,ogawa_ContinuumSurpress_prb_91_rapid}.

The absorption line shape of the 1D continuum states was recently measured~\cite{akiyama_PLEandcontinuum_apl_03} in a photoluminescence-excitation
(PLE) spectrum of T-shaped quantum wires (T-wires) with improved homogeneity, which were fabricated by cleaved-edge overgrowth with growth-interrupt
annealing~\cite{yoshita_Flat110byAnnealing_01} in molecular-beam epitaxy. The results did indeed show reduced absorption of the continuum states, and
qualitatively support a simple model calculation~\cite{ogawa_ContinuumSurpress_prb_91_rapid,ogawa_ContinuumSurpress_prb_91}. On the other hand, some
excited-exciton states that were predicted in a more detailed calculation~\cite{szymanska_Exciton_withCoulomb_prb_01} for the T-wire geometry were not found in
the observed spectrum. This seems to indicate an inconsistency between the experiment and the theory concerning the above fundamental effect,
which must be solved before this effect can be applied to optical devices such as quantum wire
lasers~\cite{wegscheider_firstLasing_93,wegscheider_currentInjection_94,rubio_CoexistenceLasing_01,hayamizu_LasingSingleWire_02}.  

In this work, we demonstrate PLE spectroscopy of highly uniform T-wires in an improved optical configuration that allows us to measure
polarization-dependent PLE spectra with less scattering-light noise than before~\cite{akiyama_PLEold_96,akiyama_PLEandcontinuum_apl_03}. The observed
spectra of the T-wires exhibit small isolated peaks, found to be in good agreement with the excited-exciton states predicted in the
calculation~\cite{szymanska_Exciton_withCoulomb_prb_01}. The peak assignments are supported by the polarization-dependence of the PLE spectra.  

\begin{figure}
\includegraphics{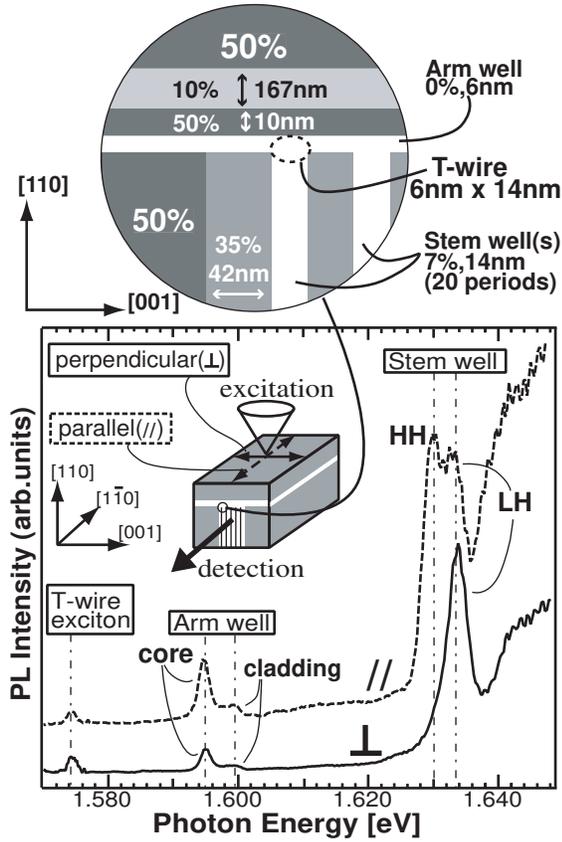}%
\caption{\label{fig1}Schematic of a T-wire-laser sample, optical configuration, and PLE spectra of T-wires at 5 K at an input power of
 6.7 $\mu$W. Percentages show Al-concentration $x$ in  ${\rm Al}_{x}{\rm Ga}_{1-x}{\rm As}$. 
PLE was measured via point spectroscopy with a spot size about 0.8 $\mu$m in diameter. The
 polarization of the light used for excitation in PLE was perpendicular ($\bot$, normal line) or parallel (//, broken line) to the T-wires.}
\end{figure}

We studied several samples that had a nominally identical T-wire-laser structure containing 20 T-wires formed at 20 T-shaped
intersections of 20 periods of 14 nm ${\rm Al}_{0.07}{\rm Ga}_{0.93}{\rm As}$ quantum wells (stem wells) and a 6 nm GaAs quantum well (arm well),
as schematically shown in Fig. 1. The barrier layers in the core and cladding regions had different Al concentrations, 35\% and
50\%, respectively, so the arm wells in the core and cladding regions had slightly different energies. 
Details of the structure and fabrication method are presented elsewhere~\cite{akiyama_PLEandcontinuum_apl_03}. 
The configuration for optical excitation and detection is also shown in Fig. 1. The excitation was performed via a 0.5 numerical aperture objective lens
through a (110) GaAs surface using a continuous-wave titanium-sapphire laser at an input power of 6.7 $\mu$W (Fig. 1) or 2.4 $\mu$W (Fig. 2). The excitation
light had polarization perpendicular($\bot$) or parallel (//) to the T-wires. In order to eliminate intense backward scattering of the excitation light,
photoluminescence (PL) from the sample was collected via a (\={1}10) surface. 

Figure 1 shows PLE spectra for perpendicular ($\bot$) and parallel (//) polarizations at 5 K. Large peaks denoted ``HH'' and ``LH'' are due to 2D
heavy-hole and light-hole excitons in the stem wells, and they demonstrate the typical polarization anisotropy of (001) GaAs quantum wells. The two peaks
denoted ``core'' and ``cladding'' are due to 2D heavy-hole excitons in the core arm well on 35\% barrier and the cladding arm well on 50\% barrier,
respectively, which were confirmed by a separate PL imaging experiment. In those peaks, optical anisotropy exists due to the crystallographic
anisotropy of a (110) GaAs quantum well~\cite{Gershoni_Anisotropy110_prb_91,Akiyama_polarizationPLE_prb_96}. The lowest energy peak is due to the 1D
ground-state excitons in the T-wires. 
The PLE spectrum for the parallel polarization reproduces our previously reported spectrum~\cite{akiyama_PLEandcontinuum_apl_03}. 

\begin{figure}
\includegraphics{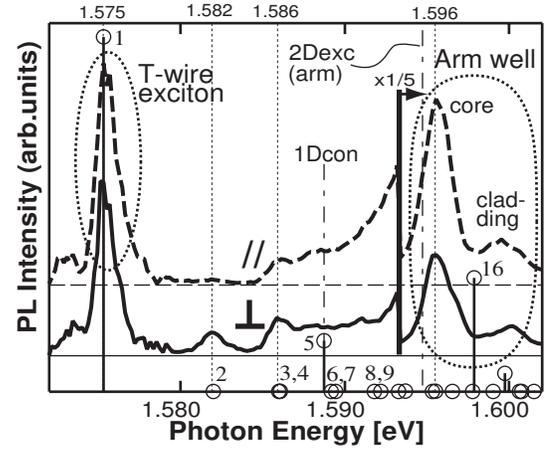}%
\caption{PLE spectra of T-wires at 5 K at an input power of 2.4 $\mu$W for perpendicular ($\bot$, normal line) and parallel (//, broken line)
 polarizations, with calculated oscillator strength (circles).
The circles are labeled by index $n$ of the states.
The dashed line represents the calculated onset of the 1D continuum state and the peak of the 2D ground-state exciton.
Horizontal lines (solid and broken) represent the zero-lines of the PLE spectra.}
\end{figure}

Figure 2 shows PLE spectra for perpendicular ($\bot$) and parallel (//) polarizations at 5 K measured in detail around the low energy region in Fig. 1. 
The strong peak at the lowest energy (1.575 eV) shown in both the spectra is due to 1D ground-state excitons in the T-wires. Between this peak and the 2D
arm-well exciton peak, continuous absorption with an onset at 1.586 eV is observed for parallel polarization, while small peaks are observed at 1.582
and 1.586 eV (7 and 11 meV above the ground-state-exciton peak) in perpendicular polarization together with the continuous absorption above
1.586 eV. The origins of the small peaks is most likely the exciton states formed with the electron ground-subband and hole excited-subbands, for the peaks
showing strong polarization dependence. The continuous absorption band is expected to correspond to the 1D continuum states. A detailed discussion
will now be given comparing a data with a theoretical calculation. 

We performed a model calculation of a single electron-hole pair, or an exciton, in the present 14 nm$\times$6 nm T-shaped geometry  using a
formulation reported previously~\cite{szymanska_Exciton_withCoulomb_prb_01}. 
Our model assumes a single hole band corresponding to the ground hole
subband in the arm well, which has an anisotropic effective mass due to
 the band-mixing effect~\cite{Sercel_APL_90,Sercel_PRB_90,brinkmann_rydbelg_prb_97,Kyrychenko_PRB_00,Watanabe_JJAP_02}.
Additional band-mixing effects due to lateral confinement is neglected,
since the second hole subband in the arm well is deviated
over 30 meV.
Calculated results are accurate in the low energy region
of our interest.
The calculation is based on exact diagonalizations of Coulomb
interaction matrix elements, which gives energy eigenvalues $E_n$ and electron-hole envelope wavefunctions
$\Psi_{n}(x_e, y_e, x_h, y_h, z=z_e-z_h)$ $(n=1,2,3, ...)$ of not only the ground but also the excited states of the exciton system. Here,
$( x_e, y_e, z_e)$ and $( x_h, y_h, z_h)$ denote the positions of an electron and a hole.
The method of the calculation is the conjugate-gradient minimization~\cite{Payne_CG_RMP_92} in
a coordinate where $x_e, y_e, x_h$ and $y_h$ move within 42 nm length,
and $z$ goes from -50 nm to 50 nm. Since we use a dense grid, the results are well converged.
 Though results for low-energy confined states are accurate, care must be taken over results for high-energy
states which are not accurate, because of the extension in the finite
sized box and the mixing effect of high energy bands. Thus, the energies of 2D exciton states in the arm well and 1D continuum states in the
T-wires were calculated by another method. 

The oscillator strength of the $n$-th state was evaluated as $| \int\int dx dy \Psi_n(x, y, x, y, 0) |^2$, a square of an integral of an electron-hole
wavefunction with respect to the same electron and hole positions. This evaluation is justified for wavefunctions spatially smaller than the
wavelength of light in conventional far-field optical detections. In a previous paper~\cite{szymanska_Exciton_withCoulomb_prb_01}, oscillator strength
was evaluated as $\int\int dx dy | \Psi_n(x, y, x, y, 0) |^2$, which is applicable for a particular detection via a local probe in a near-field scanning
optical microscope. Such ``local-probe oscillator strength'' is not used in the present analysis.  

For comparison with experimental data, we added an offset energy of 1.521 eV (corresponding to the bulk GaAs band-gap energy) to $E_n$, and plotted the
calculated oscillator strengths of the $n$-th states by circles in Fig. 2. The energies of the 1D continuum states and the arm-well
2D-exciton states are shown by dashed lines.  
The calculated energy of the $n$=1 state with large oscillator strength and the 2D exciton energy in the arm well both show good agreement with the
measured strong PLE peaks of the T-wire excitons at 1.575 eV and the arm well excitons at 1.596 eV. 
Though the oscillator strength of the $n$=3 state is zero, those of the $n$=2 and 4 states have finite small values. The calculated energy
positions as well as the small oscillator strength of the $n$=2 and 4 states show good agreement with the measured small PLE peaks at 1.582 and
1.586 eV.  

\begin{figure*}
\includegraphics{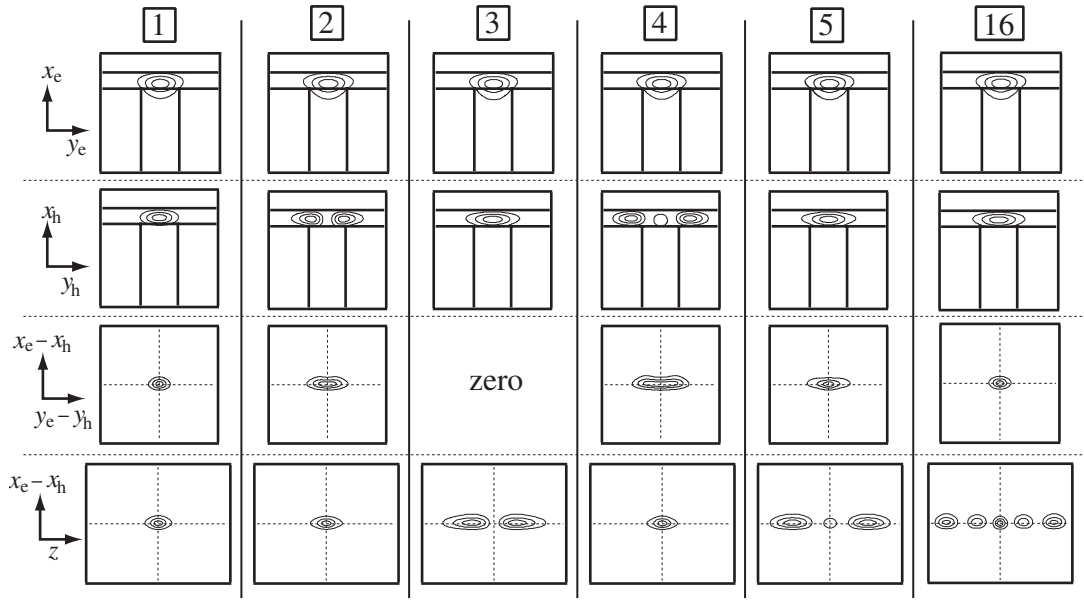}%
\caption{Contour-plots of calculated $|\Psi_n|^2$ for $n$=1, 2, 3, 4, 5 and 16 for various parameter pairs. Parameters ($x_e,y_e,z_e$) and ($x_h,y_h,z_h$)
 denote the positions of an electron and a hole, and $z$ is defined by $z = z_e - z_h$. The directions of $x$,$y$ and $z$ correspond to
 [110], [001], and [\={1}10]. The size of the boxes is 42 nm $\times$ 42 nm for the upper two rows, 84 nm $\times$ 84 nm for the third row, and
 84 nm $\times$ 100 nm for the bottom row. The three contour lines in each box represent 80\%, 50\%, and 20\%
 lines of maximum. In the relative $xy$ coordinate system, since $|\Psi_n|^2$ is plotted for z relative = 0, it is zero everywhere in the $n$=3 state which
 has a node at $z = 0$. } 
\end{figure*}

Figure 3 shows wavefunctions $\Psi_n(x_e, y_e, x_h, y_h, z=z_e-z_h)$ for $n$= 1-5 and 16, where $|\Psi_n|^2$ for various parameter pairs are plotted
in the way described in the previous paper~\cite{szymanska_Exciton_withCoulomb_prb_01}. 
The wavefunction for the $n$=1 state has no node against any parameter axis. It is well confined in the T-wire and demonstrates that the $n$=1 state is
the ground-state exciton in the T-wire. This confirms that the observed strong PLE peak at 1.575 eV is due to the ground-state exciton in the T-wire.

For $n$=2 and 4, wavefunctions have no node in the electron $(x_e, y_e)$ motion and the electron-hole relative $z$ motion, while nodes appear in the
hole $y_h$ motion, which show that the $n$=2 and 4 states are exciton states consisting of the electron ground subband and the hole excited
subbands. Errors in the calculated energy eigenvalues of these excited states in the finite-sized-box are small, because of the heavy effective mass
of hole. The oscillator strengths of these states are small, but not zero, because wavefunctions for different subbands of electrons and holes are approximately
orthogonal. The calculated energy and oscillator strength of these states agree well with the experimentally observed two small peaks at 1.582 and 1.586 eV.

Since the present calculation assumes a single hole band, it does not evaluate absolute values of polarization-dependent absorption intensities. However,
different hole subbands in general cause different polarization-dependence. Thus, different hole subbands contained in the $n$=1, 2, and 4 states
qualitatively explain the experimental finding that the two peaks at 1.582 and 1.586 eV show stronger absorption for perpendicular polarization,
unlike the ground-state exciton peak.  

For $n$=3, 5, and 16, wavefunctions have no node for the electron $(x_e, y_e)$ motion and the hole $(x_h, y_h)$ motion, while nodes appear in the
electron-hole relative $z$ motion, which show these states are 1D-exciton excited states consisting of the electron ground
subband and the hole ground subband in the T-wire. The $n$=3 wavefunction is an odd function of $z$ and has vanishing oscillator strength, whereas the
$n$=5 and 16 wavefunctions are even functions of $z$ and have large oscillator strength.
We should note here that all these states have extended wavefunctions for $z$, for which the calculated oscillator strength and energy eigenvalues are
not accurate because of the finite-sized-box and the light effective mass of electron, and cannot be compared directly with the experimental
spectrum. In fact, the separately calculated onset of 1D continuum states is at 1.589 eV, and all the exciton bound states including the $n$=5 and 16
states should be below this energy.  

An important point we learn for these states is that these 1D-exciton excited states with even parity consisting of the electron ground subband and
the hole ground subband in the T-wire have dominant oscillator strength, while other excited states have only negligible oscillator strength. On the
basis of this qualitative point, the experimentally observed continuous absorption band around 1.589 eV is ascribed to excited exciton states and 1D
continuum states consisting of the electron ground subband and the hole ground subband in the T-wire. The similarity in polarization dependence
between the continuous absorption band and exciton ground state suggests that the contributing hole subband is common.  

In summary, we measured polarization-dependent PLE spectra of highly uniform T-wires, and found good agreements with a model calculation. The lowest
energy peak due to the 1D-exciton ground state and a continuous absorption band due to 1D continuum states show polarization dependence similar to
each other, while
two small peaks due to excited hole subbands have different polarization anisotropy. The oscillator strength of the exciton ground state in T-wires is
very large, in stark contrast with the small oscillator strength of the excited exciton states and the 1D continuum states. Inverse-square-root
singularity was absent in the absorption line shape of 1D continuum states in agreement with
theories~\cite{ogawa_ContinuumSurpress_prb_91_rapid,ogawa_ContinuumSurpress_prb_91,szymanska_Exciton_withCoulomb_prb_01}.  

This work is partly supported by a Grant-in-Aid from the MEXT, Japan.

\newpage
\bibliography{apl03march}

\end{document}